\documentclass[fleqn,usenatbib]{mnras}
\usepackage{newtxtext,newtxmath}

\usepackage[T1]{fontenc}
\usepackage{ae,aecompl}

\usepackage{graphicx}	
\usepackage{amsmath}	
\usepackage{amssymb}	
\usepackage[flushleft]{threeparttable}

\title[Study of central light concentration]{Study of central light concentration in nearby galaxies}

\author[Aswathy \& Ravikumar]{
S. Aswathy,$^{1}$\thanks{E-mail: aswathysahaj@gmail.com (AS)}
C. D. Ravikumar,$^{1}$
\\
$^{1}$Department of Physics, University of Calicut, Malappuram-673635, India\\
}

\date{Accepted XXX. Received YYY; in original form ZZZ}

\pubyear{2018}

\begin{document}
\label{firstpage}
\pagerange{\pageref{firstpage}--\pageref{lastpage}}
\maketitle

\begin{abstract}
We propose a novel technique to estimate the masses of super massive black holes (SMBHs) residing at the centres of massive galaxies in the nearby Universe using simple photometry.  Aperture photometry using SEXTRACTOR is employed to determine the central intensity ratio (CIR) at the optical centre of the galaxy image for a sample of 49 nearby galaxies with SMBH mass estimations. We find that the CIR of ellipticals and classical bulges is strongly correlated with SMBH masses whereas pseudo bulges and ongoing mergers show significant scatter. Also, the CIR of low luminosity AGNs in the sample shows significant connection with the 5 GHz nuclear radio emission suggesting a stronger link between the former and the SMBH evolution in these galaxies. In addition, it is seen that various structural and dynamical properties of the SMBH host galaxies are correlated with the CIR making the latter an important parameter in galaxy evolution studies. Finally, we propose the CIR to be an efficient and simple tool not only to distinguish classical bulges from pseudo bulges but also to estimate the mass of the central SMBH.
\end{abstract}




\section{Introduction}

Super massive black holes (SMBHs) residing at the cores of nearby massive spheroids have been occupying the central stage in galaxy evolution studies over the past few decades. The masses of these intriguing objects scale with many of the structural and dynamical properties of their host spheroids implying the possibility of a galaxy-SMBH co-evolution. 
 
The first among the host galaxy parameters to have shown a significant association with the SMBH masses ($M_{\rm smbh}$) was the bulge's luminosity \citep{KR95,MH03,LF14}. It was also reported that all early-type galaxies (ETGs) with $M_{B}$ $\leq$ -18 host a central SMBH whose mass scales linearly with the spheroid stellar mass \citep{KR95,HO99,HU08,SA11}. Among the series of scaling relations reported, the most influential was the discovery of the strong correlation between the black hole mass and the stellar velocity dispersion ($\sigma$) of the bulge component of the host galaxy \citep{FM00,GE00,TR02,GU09}. Correlations of SMBH masses with the total galaxy luminosity, the concentration index of the bulge and total number of globular clusters were also noticed \citep{GR01,GD07,BT10,BE12,SA13}.

The correlations involving $M_{\rm smbh}$ and host galaxy properties seem to depend on the nature of host galaxies. \citet{KK04} suggested the existence of classical and pseudo bulges in disc galaxies and this argument was supported by various studies over the years \citep{KC06}. Classical bulges are similar in nature to elliptical galaxies and share the same fundamental plane correlation with ellipticals. They are believed to be formed in major galaxy mergers in the same way as elliptical galaxies. Pseudo bulges are more disc-like compared to classical bulges and they might have originated from secular evolution \citep{KK04}. This distinction between the formation scenarios of the two types of bulges is supposed to be reflected in their scaling relations with $M_{\rm smbh}$. Classical bulges are correlated with $M_{\rm smbh}$ whereas pseudo bulges deviate from the correlation \citep{HU08}.

The observed correlations of SMBH masses and host galaxy properties with their negligible intrinsic scatter paved the way for an onset of theoretical studies trying to explain them. Some of them explored the possibility of an active galactic nuclei (AGN) feedback hinting at a galaxy-SMBH co-evolution \citep{SR98,KI03,FA12}. Others proposed co-habitation instead of co-evolution where the correlations are governed by the merging  sequence \citep{PE07,JM11}.

The scaling relations are being modified over the past few years owing to the rapidly advancing techniques employed in obtaining black hole demographics. In addition to the modelling of stellar kinematics, adaptive optics and integral-field spectroscopy are also being used in the estimation of SMBH masses \citep{MM15,GB17}. In this light, making use of the updated SMBH masses, we attempt a photometric characterisation of the nature of SMBH hosting galaxies in the nearby Universe by studying the central three arcsec region. The photometric studies have always been limited by the effects of point spread function (PSF), and contamination by the surrounding light in the galaxy (along the line of sight). The method devised in this study is based on the concentration of optical light at the centre of the galaxy image reducing the influence of limiting factors to a minimum.

This paper is organized as follows. Section \ref{sec:section2} describes the properties of the sample galaxies and the data reduction techniques devised in this study. Section \ref{sec:section3} deals with various correlations followed by discussion and conclusion in Section \ref{sec:section4}.

\section{THE DATA}
\label{sec:section2}
We constructed a sample of 91 galaxies from \citet{PE10}, \citet{KH13} and \citet{SAGR16} subjected to the availability of the archival images by Hubble Space Telescope ($HST$) using Wide Field Planetary Camera 2 (WFPC2) in the F814W filter and SMBH mass measurements (preferably using dynamical methods) from the literature. However, we have excluded those galaxies with defects or bad pixels in the central region, or images which did not include the complete three arcsec aperture around the galaxy centre. This reduced the sample size to 49 galaxies comprising of 30 elliptical, 13 lenticular and 6 spiral galaxies. The properties of the sample galaxies are summarised in Table \ref{tab:table1}. Spheroid masses have been adopted from \citet{KH13}. Stellar ages, total dynamical masses and half-light radii have been taken from \citet{DF16}. The nuclear radio luminosity used in the analysis comes from \citet{NY16}.

\begin{table*}
	\centering
	\caption{The table lists the properties of sample galaxies. Name of the galaxy (column 1), ellipticals with cores are marked with a star while the others are not, as per the classification given by \citet{KH13}, Distance (2), the CIR computed in F814W band (3), uncertainty in the estimation of the CIR (4), mass of the SMBH (5) and references for distance and mass of the SMBH (6), stellar velocity dispersions adopted from Hyperleda (7), half-light radii (8) and dynamical masses of the galaxies (9) adopted from \citet{DF16}, dynamical masses of the spheroid components of galaxies (10) adopted from \citet{KH13}, stellar age computed from population synthesis models (11) adopted from \citet{DF16} and morphological code (12) based on the classification by \citet{KH13} and NED. }
	\label{tab:table1}
    \begin{threeparttable}
	\begin{tabular}{llcccccccccr} 
		\hline
		 Galaxy & Dist. &CIR & $\Delta_{CIR}$  & $M_{\rm smbh}$ & Ref $^{a}$ & $\sigma$  & $R_{\rm e}$ & log $M_{\rm g}$  & log $M_{\rm b}$ & SSP age & Mor.  \\ 
		 & (Mpc) & & & [10$^{8} \rm M_{\sun}]$ & & (km s$^{-1}$)  & (pc) & (M$_{\sun}$) & (M$_{\sun}$) & (Gyr) & code $^{b}$ \\
         \hline
	IC 1459$^{*}$& 28.92 &	0.92&	0.03	&	24&	SG16	&	294	&	5097.3	&	11.67	&	11.6	&	-	&	1	\\
	IC 1481& 89.90 &	1.11&	0.08	&	0.149&	KH13	&	-	&	-	&	-	&	-	&	-	&	6	\\
	IC 2560& 40.7 &	1.44&	0.09	&	0.044&	KH13	&	-	&	9143.5	& 11.58 &	10.12	&	-	&	5	\\	
	IC 4296& 40.7 &	0.78&	0.02	&	11&	SG16	&	327	&	14189.6&	11.9	&	-	&	-	&	1	\\	
	M 31& 0.774 &	1.27&	0.03	&	1.4&	SG16	&	157	&	-	&	-	&	10.35	&	-	&	4	\\
	M 87$^{*}$& 15.6 &	0.53&	0.02	&	58&	SG16	&	323	&	6414.1	&	11.43&	11.7	&	17.7	&	1	\\	
	NGC 524& 23.3 &	0.82&	0.03	&	8.3&	SG16	&	237&	4880.3&	11.1&	11.26&	12.3	&	2	\\				
	NGC 821& 23.4 &	1.23&	0.06	&	0.39&	SG16	&	198	&	4039.6	&	10.79&	10.98	&	11	&	2	\\	
	NGC 1332& 22.3	&0.88&	0.02	&	14&	SG16	&	312	&	-	&	-	&	11.27	&	-	&	1	\\
	NGC 1399$^{*}$& 19.4 &	0.58&	0.01	&	4.7&	SG16	&	334	&	718.4	&	-	&	11.5	&	-&	1	\\	
	NGC 2748& 23.4 &	0.67&	0.05	&	0.444&	KH13	&	96	&	-	&	-	&	9.41	&	-	&5	\\	
    NGC 2778& 22.3 & 1.53  & 0.12 & 0.15 & SG16 & 154 & 2802.2 & 10.2 & 10.26 & 13.4 & 1\\
    NGC 2787& 7.3 &	0.85&	0.03	&	0.40&	SG16	&	194	&	-	&	-	&	9.78	&	-	&	5\\	
	NGC 2974& 20.9 & 1.20&	0.04	&	1.7&	SG16	&	233	&	3309.8	&	10.83&	-	&	9.3	&1	\\		
   NGC 3079& 20.7 &	0.35&	0.02	&	0.024&	SG16	&	175	&	-	&	-	&	-	&	-	&	$5^{c}$	\\
   NGC 3368& 10.62 &	0.85&	0.02	&	0.077&	KH13	&	120	&	-	&	-	&	10.26	&	-	&	5	\\ 	    NGC 3377& 10.9 &	1.41&	0.04	&	0.77&	SG16	&	137	&	2066.5	&	10.17&	10.5	&	7	&	1	\\	
	NGC 3384& 11.3 &	0.87&	0.02	&	0.17&	SG16	&	146	&	2051.4	&	10.26	&	10.34	&	7.7	&	3	\\
    NGC 3489& 11.7 &	1.34&	0.03	&	0.058&	SG16	&	105	&	1046.5	&	9.89	&	10.11	&	2.5	&	5\\	
   NGC 3607$^{*}$ & 22.2 & 0.67 & 0.02 & 1.3  & SG16 & 224 & 4713.2 & 11.04 & 11.26 & 10.3 & 6$^{d}$  \\
	NGC 3608$^{*}$& 22.3 &	1.21&	0.06	&	2.0&	SG16	&	193	&	3488.5	&	10.65&	11.01	&	9.9	&	1\\	
    NGC 3842$^{*}$ & 98.4  & 0.91 & 0.05 & 97  & SG16 & 309 &  12262.5 & 11.88 & 11.77 & - & 1\\
   NGC 3945& 19.9 &	0.86&	0.02	&	0.088&	KH13	&	182	&	3136	&	10.72	&	10.5	&	10.1	&	3	\\
   NGC 4026& 13.2 &	1.21&	0.05	&	1.8&	SG16	&	173	&	1437.8	&	10.28&	10.33	&	9.9	&	2	\\		NGC 4261$^{*}$& 30.8 & 0.69&	0.02	&	5&	SG16	&	296	&	6892.7	&	11.42&	11.65	&	16.2	&	1	\\	
	NGC 4278& 16.1 &	0.89&	0.03	&	3.39&	P10	&	234	&	2511.3&	10.78	&	-	&	11.8	&	1	\\	
	NGC 4291*& 25.5 &	1.21&	0.06	&	3.3&	SG16	&	290	&	2400.3	&	11.35&	10.85	&	-	&	1	\\
   NGC 4342 & 23 &  1.80 &  0.07 &  4.5  &  SG16 & 242  & 465.5 & 10.22 & 10.31 & 17.7 & 2\\
	NGC 4374$^{*}$& 18.51 &	0.60&	0.01	&	9.25&	KH13	&	275	&	5057	&	11.28&	11.62	&	14.9	&	1	\\	
	NGC 4382$^{*}$& 17.88 &	0.92&	0.02	&	0.130&	KH13	&	175	&	6846.2	&	11.15	&	11.51	&	6.7	&	6	\\
	NGC 4458& 17.2 &	1.63&	0.10	&	0.120&	P10	&	97	&	1835.3	&	9.73	&	-	&	12	&	1	\\
	NGC 4459& 15.7 &	1.26&	0.04	&	0.68&	SG10	&	172	&	3520.6	&	10.62&	10.88	&	7	&	1	\\	
   NGC 4486B & 16.26 & 1.72 & 0.09 & 6 & KH13  & 166  & 195.5 & 10.16 & 9.64 & 11.2 & 1 \\
	NGC 4494& 17.1 &	1.14&	0.03	&	0.550&	P10	&	149	&	3162.9	&	10.69	&	-	&	8	&	1	\\
	NGC 4526& 16.44 &	0.98&	0.03	&	4.51&	KH13	&	224	&	3022	&	10.94&	11.02	&	11	&	2	\\	
	NGC 4552& 15.4 &	0.91&	0.02	&	4.267&	P10	&	250	&	3116.9	&	10.9	&	-	&	12.6	&	1	\\
	NGC 4589& 22.0 &	1.04&	0.05	&	2.691&	P10	&	219	&	6151.5	&	11.4	&	-	&	-	&	1	\\
	NGC 4649$^{*}$& 16.46 &	0.50&	0.01	&	47.2&	SG16	&	329	&	6173.7	&	11.42&	11.64	&	17.7	&	1	\\	
	NGC 4889$^{*}$& 103.2 &	0.65&	0.03	&	210&	SG16	&	393	&	15005.3&	12.05&	12.09	&	2.8	&	1	\\		
	NGC 5018& 39.4 &	1.16&	0.03	&	2.09&	P10	&	207	&	-	&	-	&	-	&	-	&	1	\\
	NGC 5516$^{*}$& 55.3 &	1.00&	0.01	&	36.9&	KH13	&	309	&	-	&	-	&	11.6	&	-	&	1	\\
	NGC 5576& 24.8 &	1.13&	0.05	&	1.6&	SG16	&	182	&	3254.7	&	10.58&	11	&	9.1	&	1\\		
	NGC 5845& 25.2 &	1.37&	0.06	&	2.6&	SG16	&	230	&	644.7	&	10.19&	10.57	&	11.5	&	1	\\	
	NGC 5846*& 24.2 &	0.66&	0.02	&	11&	SG16	&	237	&	7801	&	11.27&	-	&	17.7	&	1\\		
	NGC 6251$^{*}$& 104.6 &	1.02&	0.05	&	5&	SG16	&	311	&	-	&	-	&	11.88	&	-	&	1	\\
	NGC 7052$^{*}$& 66.4 &	0.81&	0.04	&	3.7&	SG16	&	278	&	-	&	-	&	11.61	&	-	&	1	\\
	NGC 7457& 12.53 &	1.20&	0.07	&	0.090&	KH13	&	68	&	2342	&	9.92	&	9.56	&	3.8	&	2	\\
	NGC 7619$^{*}$& 51.5 &	1.02&	0.04	&	25&	SG16	&	316	&	8375.7	&	11.84&	11.65	&	15.4	&	1	\\	
	NGC 7768$^{*}$& 112.8 &	0.66&	0.04	&	13&	SG16	&	288	&	-	&	-	&	11.75	&	-	&	1	\\

	\hline
    \end{tabular}
    \begin{tablenotes}
      \small
      \item References. $a$: KH13- \citet{KH13}, SG16- \citet{SG16}, P10- \citet{PE10}; $b$: 1-ellipticals, 2-S0s with classical bulges, 3-S0s with pseudo bulges, 4-spirals with classical bulges, 5-spirals with pseudo bulges and 6-ongoing mergers; $c$: classification is given by \citet{DG17}; $d$: classification is from \citet{AS07}.
    \end{tablenotes}
    \end{threeparttable}
    
\end{table*}

\subsection{ Data Reduction}

We have carried out aperture photometry (MAG$\_$APER) using source extractor (SEXTRACTOR, \citealt{BA96}) for two circular apertures centred at the optical centre of the galaxy image. We have selected radii of 1.5 (R$_{1}$) and three (R$_{2}$) arcsecs for the inner and outer apertures respectively. 

The magnitudes in the two apertures are used to calculate the intensity ratio at the centre of the galaxy image.  The central intensity ratio (CIR) is defined as

\begin{equation}
   CIR = \frac{I_{1}}{I_{2} - I_{1}} = \frac{10^{0.4 (m_{2}-m_{1})}}{1-10^{0.4 (m_{2}-m_{1})}}.
	\label{eq:cir}
\end{equation}

where $I_{1}$ and $I_{2}$ are the intensities and $m_{1}$  and $m_{2}$ are the magnitudes of the light within the inner and outer apertures, respectively.

\section{Results}
\label{sec:section3}
The CIR is computed for the sample galaxies in F814W filter and the values are tabulated in Table \ref{tab:table1}. We find that the CIR is related to the properties of central SMBH and the host galaxies. We have computed linear correlation coefficients for each of the correlations listed below and these are tabulated in Table \ref{tab:table2} along with the parameters used for linear fit.
\begin{table}
	\centering
	\caption{The table lists the best-fitting parameters for the relation  x = $\alpha$ CIR + $\beta$ and correlation coefficients for various relations. N denotes the number of galaxies. }
	\label{tab:table2}
	\begin{tabular}{lcccccc} 
		\hline
		x & $\alpha$ & $\beta$ & $r$ & $p$ & N \\ 
		\hline
    	log $M_{\rm smbh}$  & -1.88 $\pm$ 0.17 & 4.36 $\pm$ 0.17 & -0.80 & > 99.99 & 33 \\
        $\sigma$ 	  & -0.38 $\pm$ 0.05 & 2.74 $\pm$ 0.05  &  -0.80 & > 99.99 & 33\\
        $M_{\rm bulge}$ & -1.41 $\pm$ 0.23 & 12.60 $\pm$ 0.23 &  -0.85 & > 99.99 & 26  \\
        $M_{\rm gal}$ & -1.49 $\pm$ 0.26 & 12.48 $\pm$ 0.27 & -0.78 & > 99.99 & 26 \\
        SSP age & -10.8 $\pm$ 1.48 & 22.77 $\pm$ 1.52& -0.88 & > 99.99 & 18  \\
		log $R_{\rm e}$  & -0.77 $\pm$ 0.17  & 4.36 $\pm$ 0.17& -0.79 & > 99.99 & 28 \\
          $L_{\rm radio}$ &  -4.01 $\pm$ 0.95 & 24.26 $\pm$ 0.95 & -0.79 & 99.86 & 13 \\
\hline
	\end{tabular}
\end{table}

\subsection {Correlation between the CIR and $M_{\rm smbh}$}
\begin{figure}
\centering
\parbox{9cm}{
\centering
\includegraphics[width=9cm]{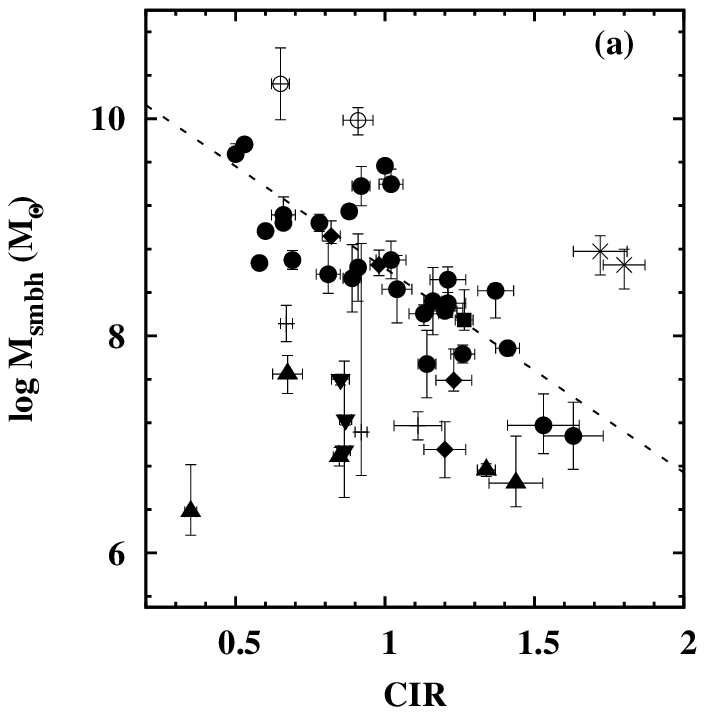}
}
\begin{minipage}{18cm}
\parbox{9cm}{
\centering
\includegraphics[width=9cm]{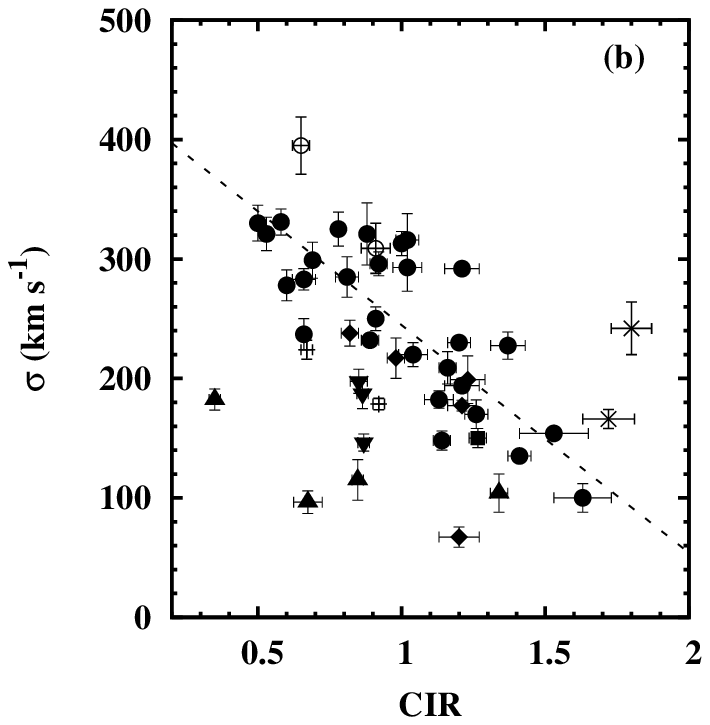}
}
\end{minipage}
\caption{ Correlation between the central intensity ratio and (a) mass of the SMBH adopted from \citet{PE10}, \citet{KH13} and \citet{SG16} (b) stellar velocity dispersion adopted from HyperLEDA database. The best-fitting line is drawn for the sub-sample consisting of ellipticals and classical bulges. Filled circles denote elliptical galaxies, filled diamonds  denote classical bulges, the square is a spiral galaxy with classical bulge, inverted triangles denote lenticular pseudo bulges, upward-pointing triangles are spiral pseudo bulges, plus marked  points are galaxies with ongoing mergers, empty  circles denote black hole monsters (reportedly over massive black holes residing in relatively small bulges) and cross marked points are tidally disrupted galaxies.}
\label{fig:figure1}
\end{figure}

We find that the CIR of all ellipticals and classical bulges in the sample is strongly correlated with $M_{\rm smbh}$, the linear correlation coefficient, $r$ being -0.80 with significance, $p$ greater than 99.99 percent  \citep{PT92}. The observed anti-correlation between the CIR and $M_{\rm smbh}$ is presented in Fig. \ref{fig:figure1}(a) along with the uncertainties involved in their respective estimations. In order to explore the effect of distance on the estimation of the CIR, we tried to calculate the CIR for a range of R$_{1}$ and found it to be varying not only with the choice of R$_{1}$, but also with the chosen S${\acute{e}}$rsic index, n. However, the variation is very minimum  when R$_{1}$ $<<$  R$_{\rm e}$ (for e.g., a variation of 0.02 when n = 3). In our sample, the  R$_{1}$/R$_{\rm e}$  for the nearest and farthest galaxy is 0.008 and 0.03, respectively. Secondly, we modified R$_{1}$(and R$_{2}$) to correspond to 1.5 (and 3) arcsecs at a distance of 16.5 Mpc (the average distance to the Virgo galaxy cluster) and redetermined the CIR. We found that the correlation coefficient did not change significantly ($r$ = -0.79, $p$ $>$ 99.99 percent) even though a minor systematic offset in the CIR was observed with distance. We further confirmed that the inclusion of the five farthest galaxies (at distances > 80 Mpc) did not affect any of the correlations significantly. Hence, we have included all galaxies in our further analysis.

It can be seen that all pseudo bulges and mergers in progress deviate from the CIR-$M_{\rm smbh}$ correlation. The only galaxy with a classical bulge which does not obey the correlation is NGC 7457. Though \citet{KH13} placed this galaxy among classical bulges, they have not ruled out the possibility of it hosting a pseudo bulge (see section \ref{sec:section4} also). Hence, the significant deviation exhibited by this galaxy might also be indicative of it hosting a dominant pseudo bulge. All other classical bulges are following the correlation between the CIR and $M_{\rm smbh}$ including the spiral galaxy M31. 

In some of the SMBH-host galaxy correlations reported in the literature, it has been shown that pseudo bulges do not correlate with the mass of the black hole residing at the centre in the same way as classical bulges and ellipticals \citep{HU08}. Our result also supports this argument. In unison with pseudo bulges, mergers in progress also are found to be outliers in Fig. \ref{fig:figure1}(a). Mergers in progress are reported to be hosting black holes with underestimated masses \citep{KH13}. It is also possible that the deviation is indicating a different formation mechanism or a different method of accretion by the central black hole.  

It might seem that the pseudo bulges and mergers in progress (denoted by triangles and plus marked points in Fig. \ref{fig:figure1}(a), respectively) excluding the edge-on spiral galaxy NGC 3379, are part of another correlation parallel to the one exhibited by ellipticals and classical bulges. This has been suggested in cases of some of the SMBH-host galaxy scaling relations \citep{KH13} but our sample size of pseudo bulges is not sufficiently large to validate it. Most of the pseudo bulges which are found to deviate significantly are barred galaxies and none of the classical bulges which obey the correlation possess bars. This might be due to the secular evolution of the barred galaxies as reported by \citet{SG16}. 

Two galaxies in our sample (NGC 4486B and NGC 4342) which are known to be tidally disrupted \citep{BF14} exhibit large offsets with respect to the CIR- $M_{\rm smbh}$ relation. Also, we have not included two black hole monsters with reportedly over-massive and possibly unreliable mass estimates as suggested by \citet{KH13} in our correlation analysis.

\subsection{Correlations between the CIR and host galaxy properties}
The CIR shows an anti-correlation ($r$ = -0.80 with $p$ > 99.99 percent)  with the stellar velocity dispersion adopted from HyperLeda database as shown in Fig.\ref{fig:figure1}(b) . In this case also, the pseudo bulges are behaving differently. The two merger galaxies in our sample with the stellar velocity measurement available also deviate from the relation. The $M_{\rm smbh}$-$\sigma$ correlation reported simultaneously by \citet{FM00} and \citet{GE00} is well known among the SMBH- host galaxy scaling relations. As the CIR is found to be strongly correlated with $M_{\rm smbh}$, it is not surprising to find that the former is correlated to stellar velocity dispersion as well. 

\begin{figure*}
\centering
\parbox{7cm}{
\includegraphics[width=7cm]{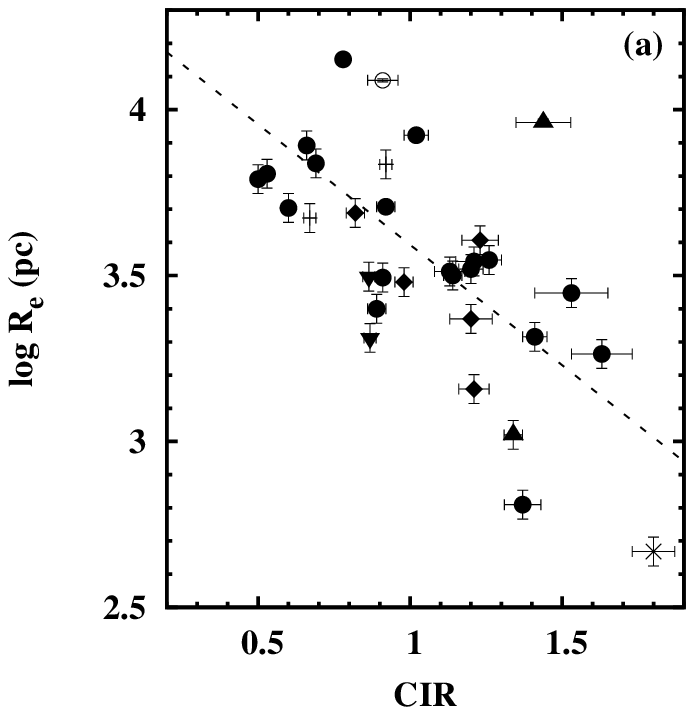}
\label{fig:figure2a}}
\qquad
\begin{minipage}{7cm}
\includegraphics[width=7cm]{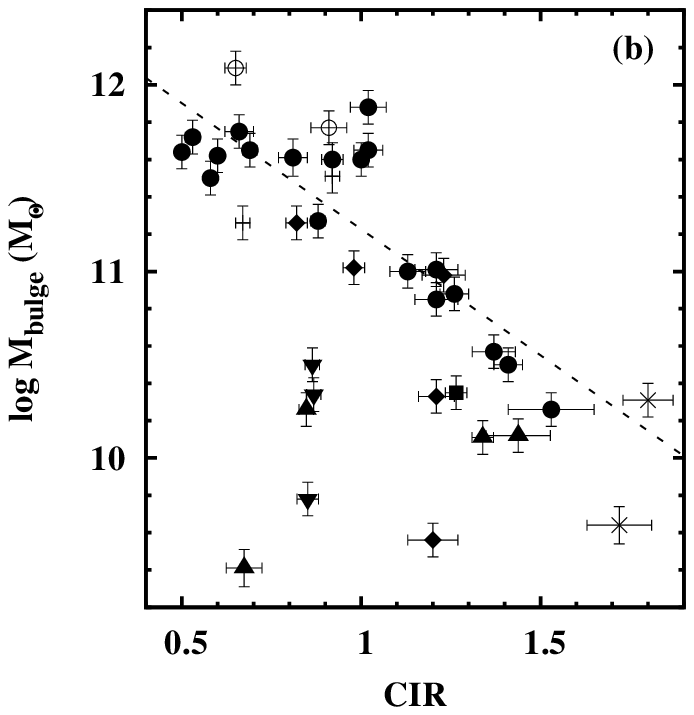}
\label{fig:figure2b}
\end{minipage} \\
\centering
\parbox{7cm}{
\includegraphics[width=7cm]{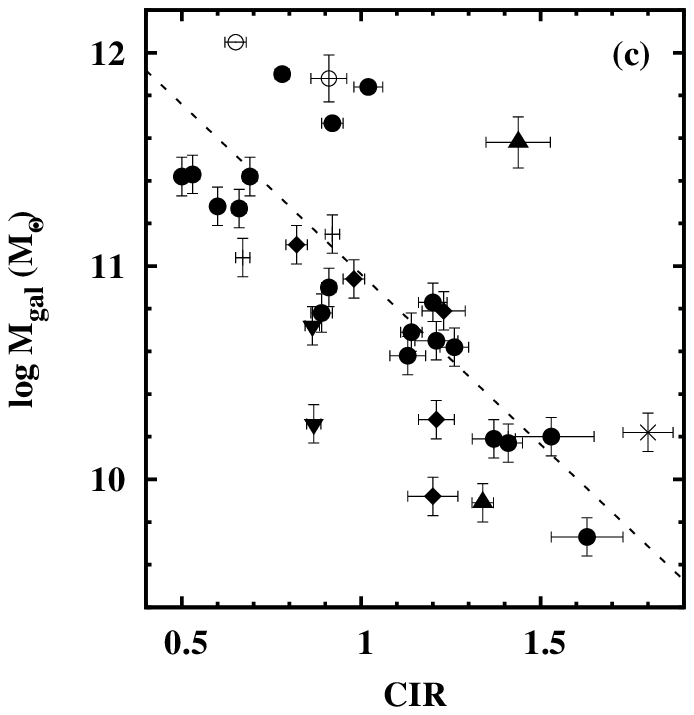}
\label{fig:figure2c}}
\qquad
\begin{minipage}{7cm}
\includegraphics[width=7cm]{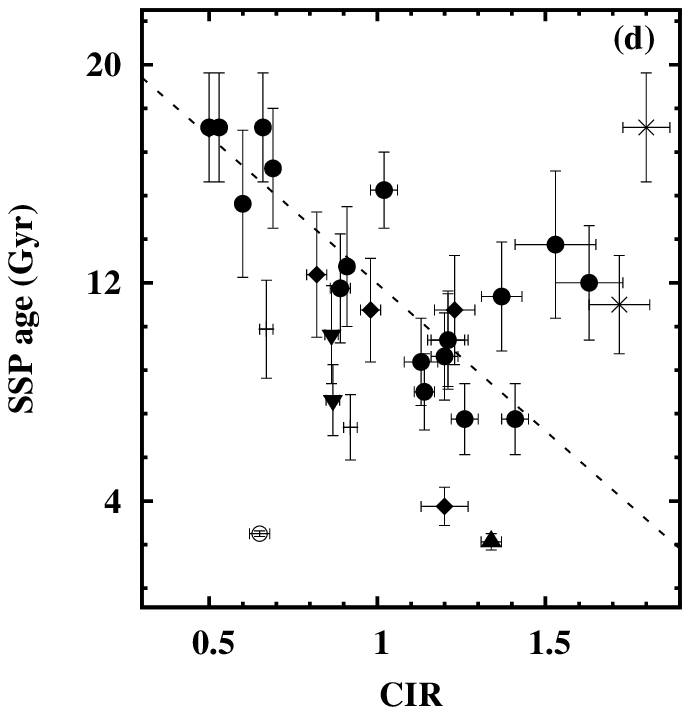}
\label{fig:figure2d}
\end{minipage} \\
\caption {Correlations between the central intensity ratio and (a)  half-light radii (R$_{\rm e}$) (b) mass of the bulge component (M$_{\rm bulge}$) (c) dynamical mass of the galaxy (M$_{\rm gal}$) and (d) age of the single stellar population where R$_{\rm e}$, M$_{\rm gal}$ and stellar ages are adopted from \citet{DF16} and M$_{\rm bulge}$ values are adopted from \citet{KH13}. The symbols denote the same objects as given in Fig. \ref{fig:figure1}}.
\label{fig:figure2}
\end{figure*}

We find that the CIR of ellipticals and classical bulges in the sample also shows an anti-correlation ($r$ = -0.79, $p$ > 99.99 percent) with the half-light radii of the galaxies adopted from the catalogue of ETGs compiled by \citet{DF16} and is shown in Fig. \ref{fig:figure2}(a). This catalogue contains $R_{\rm e}$ values estimated for different filters ranging from optical to near IR. Though this might seem to be a source of inhomogeneity, a drastic difference between $R_{\rm e}$ values estimated in $g$ and $z$ bands is not expected \citep{DF16} and we use it for statistical purpose alone, putting the observed scatter as an upper limit. Here also, the pseudo bulges show more scatter compared to ellipticals and classical bulges but the merger galaxies seem to be following the correlation.

It is seen that the CIR shows anti-correlation with the dynamical parameters of the galaxies such as the bulge mass ($M\rm_{bulge}$) and total dynamical mass ($M\rm_{gal}$) of the galaxy as shown in Fig. \ref{fig:figure2}(b) \& (c), respectively. The bulge masses have been adopted from \citet{KH13} while the galaxy masses are taken from \citet{DF16} which are estimated dynamically using  $M / L_{\rm K}$ and $M / L_{\rm V}$ values, respectively. From Table \ref{tab:table2}, we find that $M_{\rm bulge}$ is correlated better than $M_{\rm gal}$ ($r$ = -0.85 over 0.78) with the CIR. In the case of correlation of the CIR with $M_{\rm bulge}$, we can also find that pseudo bulges and classical bulges are clearly separated. This might be indicating an underlying connection between $M_{\rm bulge}$ and the CIR during the process of galaxy evolution. The merger galaxies in our sample obey both the correlations. We also notice that the CIR is correlated with bulge luminosities in $B$ and $V$ bands available from the literature but the distinction between classical bulges and pseudo-bulges is not evident in this case. 

The stellar ages estimated using population models assuming Single Stellar Population (SSP) reported by \citet{DF16} are also correlated well with the CIR ($r$ = -0.88, $p$ > 99.99 percent). Older systems are found to be possessing smaller values of the CIR as seen in Fig. \ref{fig:figure2}(d). Here also, classical bulges and pseudo bulges are clearly distinguished. The only outlier worthy of investigation is the galaxy (NGC 4889) with the highest black hole mass in the sample. \citet{CG10} reported that this galaxy is found to exhibit bimodal distribution of SSP ages. According to this study, the halo of this galaxy is uniformly old with ages between 10 and 13 Gyr whereas the inner region of this galaxy is found to contain younger population. The large offset exhibited by this galaxy with respect to the relation between the CIR and SSP age might be attributed to its bimodal stellar population.

The merger galaxies NGC 4382 and NGC 3607 show notable offsets from the fitted relation between the CIR and stellar age unlike in the cases of $R_{\rm e}$, $M\rm_{bulge}$ and $M_{\rm gal}$. Here also, we find that the pseudo bulges and mergers form a parallel relation distinct from ellipticals and classical bulges. Tidally disrupted galaxies in our sample are also not following the correlation. 

\subsection{Correlation between the CIR and central radio luminosity}
\begin{figure}
	\includegraphics[scale=1]{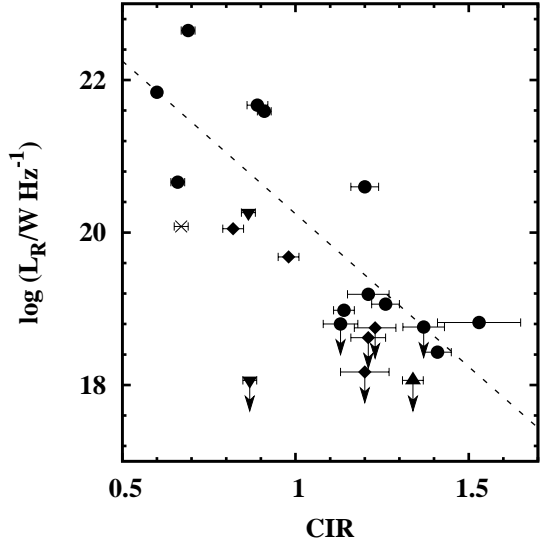}
    \caption{Correlation between the CIR and the central radio luminosity in 5 GHz band adopted from \citet{NY16}. The symbols are similar to the ones used in Fig. \ref{fig:figure1}. Downward-pointing arrows represent radio luminosity upper limits.}
    \label{fig:figure3}
\end{figure}

\citet{NY16} carried out the study of nuclear radio emission in a sub-sample of ATLAS$^{\rm 3D}$ survey of ETGs using 5 GHz, Karl G. Jansky Very Large Array. They found that more than 50 percent of their sample galaxies are low luminosity AGNs (LLAGNs). We find that 16 of our galaxies are included in this study and the luminosity of their central 5 GHz radio emission is measured. Fig. \ref{fig:figure3} shows that the CIR is strongly correlated with the central nuclear emission of these galaxies.We find that the correlation is fairly strong ( r = -0.79, p = 99.86 $\%$) for our sample excluding the radio luminosities with only upper limits. When these points are also included, the correlation co-efficient improves to -0.84 at a significance of 99.98 percent.

The relation between SMBH masses and nuclear radio emission of ETGs is not yet understood clearly. Some authors report that the two parameters are correlated \citep{NA05} whereas others dispute the correlation \citep{PA13}. The current relation between optical concentration and the central radio luminosity might be indicating a co-evolution of ETGs and LLAGNs. 

\section{DISCUSSION AND CONCLUSIONS}
\label{sec:section4}
We have performed photometric studies of the centres of SMBH hosting galaxies in the nearby Universe. This study finds that the central intensity ratio (which is a newly introduced measure of the concentration of light at the very centre of the galaxy image) of ellipticals and classical bulges correlates well with the mass of the central SMBH while spirals and lenticulars with pseudo bulges deviate from the correlation. This ratio also shows significant correlations with various parameters of the SMBH hosting galaxies such as ages and masses of the stellar population. The central intensity ratio also correlates with the central radio emission from the LLAGNs reinstating its importance in studies of galaxy evolution.

The concept of optical concentration at the centre of the galaxies correlating with the mass of the SMBH is not new. Photometric parameters such as luminosity of the spheroid and stellar concentration index \citep{KR95,GR01} were found to be correlated with the central SMBH masses. A series of papers were published based on the idea of concentration index which was first presented by \citet{TR01}. Concentration index is defined as the ratio of flux inside some fraction $\alpha$ of half-light radius to the total flux with in the half-light radius. This index was used by \citet{GR01} to obtain a correlation between galaxy light concentration and super massive black hole mass. \citet{GR03} reported a strong correlation between the shape of a bulge's light profile and the mass of its central SMBH. Such studies, however, require detailed modelling of various components of the galaxies to extract their structural and photometric properties. This might require more detailed bulge-disc decomposition algorithms, which could raise the associated uncertainty levels owing to the complexity of the method. Further, the simple method used in the present study also shows similar strengths in correlations.

\citet{GTC01} while investigating the stability of stellar concentration index devised in their study, speculated that the mass of the central black hole might be regulated by the way mass is distributed in a galaxy. The radii which are used in estimating the central intensity ratio in our study might not have a physical motivation. Yet, this concept seems to be significant as it appears to quantify the distribution of matter in the central region which is closely correlated with the $M_{\rm smbh}$ validating the hypothesis of \citet{GTC01}.

Some of the galaxies in which the disc component remains embedded in the spheroid component and does not dominate the galaxy light were found to be showing large scatter in $M_{\rm smbh}$-$M_{\rm bulge}$ correlations. Recently \citet{SAGR16} showed that some of these galaxies with intermediate-scale discs (as they are intermediate between disc-less ellipticals and disc-dominated lenticulars) have been subjected to incorrect decomposition methods. These galaxies when correctly modelled followed the correlation. The present study includes two such galaxies (NGC 1332 and NGC 4291) and these galaxies follow the observed correlation between the CIR and SMBH mass, establishing further the usefulness and significance of determination of the CIR.

We find that while ellipticals and classical bulges follow the correlation between the CIR and $M_{\rm smbh}$, pseudo bulges seem to be deviating from the correlation. The only spiral in the sample hosting a classical bulge shows the correlation whereas all other spirals hosting pseudo bulges are showing significant deviation from the observed correlation. The galaxies with ongoing mergers (NGC 3607,NGC 4382 and IC 1481) also show significant deviation.

In the case of lenticular galaxies also all classical bulges are forming an integral part of the correlation except the galaxy NGC 7457. Even though the galaxy NGC 7457 was classified to be hosting a classical bulge by \citet{KH13}, they have also discussed the possibility of it being a pseudo bulge dominant galaxy. NGC 7457 is an outlier in the correlation between the mass of the SMBH and number of globular cluster systems present in the galaxy ($M_{\rm smbh}$- $N_{\rm GC}$) reported by \citet{HH11}. This galaxy hosts a comparatively smaller black hole and it also has a remarkably low stellar velocity dispersion. The deviation of this particular galaxy from the observed correlation might be attributed to these reasons. Our result supports the argument that NGC 7457 might be a pseudo bulge dominant galaxy. Also, this indicates that the CIR can serve as a simple and powerful tool to distinguish between classical bulges and pseudo bulges using photometry alone. 

The fact that classical and pseudo bulges do not correlate in the same way with the mass of the super massive black hole is already established \citep{HU08,HU09}. The two types of bulges seem to follow different processes of evolution which make them differ in their scaling relations. Classical bulges are believed to be formed after major mergers. Pseudo bulges are formed secularly out of their discs \citep{KK04}. The pseudo bulges in our sample show significant offsets from all fitted relations and our result validates this theory. 

The elliptical galaxies are classified further based on their physical properties and evolution mechanisms as core-ellipticals and core-less ellipticals \citep{KO09}. Core ellipticals are believed to be formed as a result of dry mergers and these possess cuspy profiles near their centres. Core-less ellipticals are known to be formed through mergers involving star bursts at the centre thereby hosting young stellar population compared to core-galaxies \citep{MH94}. Classical bulges are similar to core-less galaxies and these are generally observed to be radio-quiet \citep{KH13}. Our sample includes core ellipticals, core-less ellipticals and classical bulges. The CIR shows anti-correlation with the mass of the galaxy and it decreases with increasing age of the stellar population (Fig. \ref{fig:figure2}(d)). The CIR is also anti-correlated to half light radii. We observe that all core-ellipticals in our sample with stellar ages above 14 Gyrs (except the tidally disrupted galaxy NGC 4342) have lower central intensity ratios compared to core-less galaxies such as NGC 3377 and NGC 4459. These galaxies are also radio-loud. The decrease in the central intensity ratio of these galaxies might be attributed to the dry mergers happened in the past. On the other hand, most of the classical bulges and core-less galaxies show central intensity ratios close to or more than unity. This might be due to the extra light at the centre resulting from star-bursts or wet mergers. These galaxies also contain young stellar population compared to core galaxies and many of these systems have only an upper limit to radio luminosities as seen in Fig. \ref{fig:figure3}. Hence, the CIR might also be an indicator of the evolutionary path of the galaxies.

Since the CIR-stellar age correlation is very strong and similar in nature to the CIR-$M_{\rm smbh}$ relation, the role of stellar age in establishing the CIR-$M_{\rm smbh}$ correlation needs to be explored further. However, the large uncertainties present in the data coupled with a small sample size hampers our efforts to find a plane relation between the CIR, $M_{\rm smbh}$ and stellar age. It is also possible that the ageing population alone cannot produce the observed correlation between the CIR and SMBH masses as other factors like accretion processes by the central black hole might be contributing towards it.

We find that the central 5 GHz radio emission of LLAGNs in our sample given by \citet{NY16} correlates with the central intensity ratio. The origin of this nuclear radio emission is not well understood though its origin may be linked to the synchrotron emission from the central SMBH. Another possibility proposed is the low-level circum-nuclear star formation triggers the synchrotron emission \citep{CO92}. \citet{NY16} also reported a correlation between the central radio emission and SMBH mass. According to their study, the most powerful radio sources reside in ETGS with the most massive black holes. They have also discussed the possibility of the LLAGNs possessing radiatively inefficient accretion mechanisms.  Since in most of the correlations listed in Section \ref{sec:section3}, the CIR is acting as a proxy to the black hole mass, the central radio emission might also be originating from the SMBH. Since we find that the central intensity decreases with increasing mass of the galaxy, it is quite unlikely that recent star formation occurs at the centre \citep{IM18}. Many of these systems contain old stellar population which again does not support this theory. Further, more massive galaxies, in general, have large half light radii causing their centres to be less dense as seen in \citet{MH11}. This can also result in a decrease in the central intensity ratio.

In the case of radiatively inefficient accretion as seen in ETGs, the observed scaling relations may be attributed to feedback mechanisms related to radio outflows \citep{HB14}. AGN feedback might be carried out in the form of radiative winds from energetic quasars and radio jets in LLAGNs \citep{CI10}. This feedback is capable of expelling gas from the host galaxy thereby suppressing future star formation \citep{MF13}. The turbulent energy thrown out into the ambient inter stellar medium (ISM) in the form of powerful radio jets can also prevent star formation \citep{AL15}. This scenario is supported by our result as we find low values of the central intensity ratio in more massive SMBHs. It could be that the feedback mechanism is active suppressing the star formation near the central region resulting in a decrease in the CIR. 

In most of the correlations listed above, the CIR seems to be on equal footing with the SMBH mass and provides a robust estimation of the SMBH masses. In this light, the advantages of using the CIR to predict SMBH masses might be highlighted. The correlations obtained using the CIR are fairly independent of the distance measurements in the nearby Universe and the central intensities. Also, simple Monte-Carlo simulations suggest that the CIR is practically independent of the viewing angle and fairly stable for relatively big ellipsoidal galaxies. This method is not expensive as we rely on photometric images for its computation. Most importantly, it is extremely simple to calculate as the procedure does not require any decomposition or modelling of the galaxy profiles and hence devoid of associated uncertainties. Also, the CIR does not depend significantly upon exposure depth of the images. Since we are dealing with ratios, redshift dependent corrections are not significant, at least in the nearby Universe. Our relation may be highly significant in case of distant galaxies as the spectroscopic measurements are difficult at high redshifts.

\section*{Acknowledgements}

We thank the anonymous reviewer for his/her valuable comments which greatly improved the contents of this paper. SA would like to acknowledge the financial support from Kerala State Council for Science, Technology and Environment (KSCSTE). We acknowledge the use of the NASA Extragalactic Database (NED), \href{https://ned.ipac.caltech.edu/}{https:
//ned.ipac.caltech.edu/} operated by the Jet Propulsion Laboratory, California Institute of Technology, and the Hyperleda database, \href {http://leda.univ-lyon1.fr/}{http://leda.univ-lyon1.fr/}. Some of the data presented in this paper were obtained from the Mikulski Archive for Space Telescopes (MAST), \href{http://archive.stsci.edu/}{http://archive.stsci.edu/}



\bibliographystyle{mnras}
\bibliography{CIR} 

\begin{thebibliography}{}
\makeatletter
\relax
\def\mn@urlcharsother{\let\do\@makeother \do\$\do\&\do\#\do\^\do\_\do\%\do\~}
\def\mn@doi{\begingroup\mn@urlcharsother \@ifnextchar [ {\mn@doi@}
  {\mn@doi@[]}}
\def\mn@doi@[#1]#2{\def\@tempa{#1}\ifx\@tempa\@empty \href
  {http://dx.doi.org/#2} {doi:#2}\else \href {http://dx.doi.org/#2} {#1}\fi
  \endgroup}
\def\mn@eprint#1#2{\mn@eprint@#1:#2::\@nil}
\def\mn@eprint@arXiv#1{\href {http://arxiv.org/abs/#1} {{\tt arXiv:#1}}}
\def\mn@eprint@dblp#1{\href {http://dblp.uni-trier.de/rec/bibtex/#1.xml}
  {dblp:#1}}
\def\mn@eprint@#1:#2:#3:#4\@nil{\def\@tempa {#1}\def\@tempb {#2}\def\@tempc
  {#3}\ifx \@tempc \@empty \let \@tempc \@tempb \let \@tempb \@tempa \fi \ifx
  \@tempb \@empty \def\@tempb {arXiv}\fi \@ifundefined
  {mn@eprint@\@tempb}{\@tempb:\@tempc}{\expandafter \expandafter \csname
  mn@eprint@\@tempb\endcsname \expandafter{\@tempc}}}

\bibitem[\protect\citeauthoryear{{Afanasiev} \& {Silchenko}}{{Afanasiev} \&
  {Silchenko}}{2007}]{AS07}
{Afanasiev} V.~L.,  {Silchenko} O.~K.,  2007, \mn@doi [Astronomical and
  Astrophysical Transactions] {10.1080/10556790701553524}, \href
  {http://adsabs.harvard.edu/abs/2007A%26AT...26..311A} {26, 311}

\bibitem[\protect\citeauthoryear{{Alatalo} et~al.,}{{Alatalo}
  et~al.}{2015}]{AL15}
{Alatalo} K.,  et~al., 2015, \mn@doi [\apj] {10.1088/0004-637X/798/1/31}, \href
  {http://adsabs.harvard.edu/abs/2015ApJ...798...31A} {798, 31}

\bibitem[\protect\citeauthoryear{{Beifiori}, {Courteau}, {Corsini}  \&
  {Zhu}}{{Beifiori} et~al.}{2012}]{BE12}
{Beifiori} A.,  {Courteau} S.,  {Corsini} E.~M.,   {Zhu} Y.,  2012, \mn@doi
  [\mnras] {10.1111/j.1365-2966.2011.19903.x}, \href
  {http://cdsads.u-strasbg.fr/abs/2012MNRAS.419.2497B} {419, 2497}

\bibitem[\protect\citeauthoryear{{Bertin} \& {Arnouts}}{{Bertin} \&
  {Arnouts}}{1996}]{BA96}
{Bertin} E.,  {Arnouts} S.,  1996, \mn@doi [\aaps] {10.1051/aas:1996164}, \href
  {http://cdsads.u-strasbg.fr/abs/1996A%26AS..117..393B} {117, 393}

\bibitem[\protect\citeauthoryear{{Blom}, {Forbes}, {Foster}, {Romanowsky}  \&
  {Brodie}}{{Blom} et~al.}{2014}]{BF14}
{Blom} C.,  {Forbes} D.~A.,  {Foster} C.,  {Romanowsky} A.~J.,   {Brodie}
  J.~P.,  2014, \mn@doi [\mnras] {10.1093/mnras/stu095}, \href
  {http://cdsads.u-strasbg.fr/abs/2014MNRAS.439.2420B} {439, 2420}

\bibitem[\protect\citeauthoryear{{Burkert} \& {Tremaine}}{{Burkert} \&
  {Tremaine}}{2010}]{BT10}
{Burkert} A.,  {Tremaine} S.,  2010, \mn@doi [\apj]
  {10.1088/0004-637X/720/1/516}, \href
  {http://cdsads.u-strasbg.fr/abs/2010ApJ...720..516B} {720, 516}

\bibitem[\protect\citeauthoryear{{Ciotti}, {Ostriker}  \& {Proga}}{{Ciotti}
  et~al.}{2010}]{CI10}
{Ciotti} L.,  {Ostriker} J.~P.,   {Proga} D.,  2010, \mn@doi [\apj]
  {10.1088/0004-637X/717/2/708}, \href
  {http://cdsads.u-strasbg.fr/abs/2010ApJ...717..708C} {717, 708}

\bibitem[\protect\citeauthoryear{{Coccato}, {Gerhard}  \&
  {Arnaboldi}}{{Coccato} et~al.}{2010}]{CG10}
{Coccato} L.,  {Gerhard} O.,   {Arnaboldi} M.,  2010, \mn@doi [\mnras]
  {10.1111/j.1745-3933.2010.00897.x}, \href
  {http://adsabs.harvard.edu/abs/2010MNRAS.407L..26C} {407, L26}

\bibitem[\protect\citeauthoryear{{Condon}}{{Condon}}{1992}]{CO92}
{Condon} J.~J.,  1992, \mn@doi [\araa] {10.1146/annurev.aa.30.090192.003043},
  \href {http://cdsads.u-strasbg.fr/abs/1992ARA%26A..30..575C} {30, 575}

\bibitem[\protect\citeauthoryear{{Dabringhausen} \&
  {Fellhauer}}{{Dabringhausen} \& {Fellhauer}}{2016}]{DF16}
{Dabringhausen} J.,  {Fellhauer} M.,  2016, \mn@doi [\mnras]
  {10.1093/mnras/stw1248}, \href
  {http://cdsads.u-strasbg.fr/abs/2016MNRAS.460.4492D} {460, 4492}

\bibitem[\protect\citeauthoryear{Davis, Graham  \& Seigar}{Davis
  et~al.}{2017}]{DG17}
Davis B.~L.,  Graham A.~W.,   Seigar M.~S.,  2017, \mn@doi [\mnras]
  {10.1093/mnras/stx1794}, \href {http://dx.doi.org/10.1093/mnras/stx1794}
  {471, 2187}

\bibitem[\protect\citeauthoryear{{Fabian}}{{Fabian}}{2012}]{FA12}
{Fabian} A.~C.,  2012, \mn@doi [\araa] {10.1146/annurev-astro-081811-125521},
  \href {http://cdsads.u-strasbg.fr/abs/2012ARA%26A..50..455F} {50, 455}

\bibitem[\protect\citeauthoryear{{Ferrarese} \& {Merritt}}{{Ferrarese} \&
  {Merritt}}{2000}]{FM00}
{Ferrarese} L.,  {Merritt} D.,  2000, \mn@doi [\apjl] {10.1086/312838}, \href
  {http://cdsads.u-strasbg.fr/abs/2000ApJ...539L...9F} {539, L9}

\bibitem[\protect\citeauthoryear{{Gao} et~al.,}{{Gao} et~al.}{2017}]{GB17}
{Gao} F.,  et~al., 2017, \mn@doi [\apj] {10.3847/1538-4357/834/1/52}, \href
  {http://cdsads.u-strasbg.fr/abs/2017ApJ...834...52G} {834, 52}

\bibitem[\protect\citeauthoryear{{Gebhardt} et~al.,}{{Gebhardt}
  et~al.}{2000}]{GE00}
{Gebhardt} K.,  et~al., 2000, \mn@doi [\apjl] {10.1086/312840}, \href
  {http://cdsads.u-strasbg.fr/abs/2000ApJ...539L..13G} {539, L13}

\bibitem[\protect\citeauthoryear{{Graham} \& {Driver}}{{Graham} \&
  {Driver}}{2007}]{GD07}
{Graham} A.~W.,  {Driver} S.~P.,  2007, \mn@doi [\apj] {10.1086/509758}, \href
  {http://cdsads.u-strasbg.fr/abs/2007ApJ...655...77G} {655, 77}

\bibitem[\protect\citeauthoryear{{Graham}, {Trujillo}  \& {Caon}}{{Graham}
  et~al.}{2001a}]{GTC01}
{Graham} A.~W.,  {Trujillo} I.,   {Caon} N.,  2001a, \mn@doi [\aj]
  {10.1086/323090}, \href {http://cdsads.u-strasbg.fr/abs/2001AJ....122.1707G}
  {122, 1707}

\bibitem[\protect\citeauthoryear{{Graham}, {Erwin}, {Caon}  \&
  {Trujillo}}{{Graham} et~al.}{2001b}]{GR01}
{Graham} A.~W.,  {Erwin} P.,  {Caon} N.,   {Trujillo} I.,  2001b, \mn@doi
  [\apjl] {10.1086/338500}, \href
  {http://cdsads.u-strasbg.fr/abs/2001ApJ...563L..11G} {563, L11}

\bibitem[\protect\citeauthoryear{{Graham}, {Erwin}, {Caon}  \&
  {Trujillo}}{{Graham} et~al.}{2003}]{GR03}
{Graham} A.~W.,  {Erwin} P.,  {Caon} N.,   {Trujillo} I.,  2003, in
  {Avila-Reese} V.,  {Firmani} C.,  {Frenk} C.~S.,   {Allen} C.,  eds,  Revista
  Mexicana de Astronomia y Astrofisica Conference Series Vol. 17, Revista
  Mexicana de Astronomia y Astrofisica Conference Series. pp 196--197
  (\mn@eprint {} {astro-ph/0206248})

\bibitem[\protect\citeauthoryear{{G{\"u}ltekin} et~al.,}{{G{\"u}ltekin}
  et~al.}{2009}]{GU09}
{G{\"u}ltekin} K.,  et~al., 2009, \mn@doi [\apj] {10.1088/0004-637X/698/1/198},
  \href {http://cdsads.u-strasbg.fr/abs/2009ApJ...698..198G} {698, 198}

\bibitem[\protect\citeauthoryear{{Harris} \& {Harris}}{{Harris} \&
  {Harris}}{2011}]{HH11}
{Harris} G.~L.~H.,  {Harris} W.~E.,  2011, \mn@doi [\mnras]
  {10.1111/j.1365-2966.2010.17606.x}, \href
  {http://cdsads.u-strasbg.fr/abs/2011MNRAS.410.2347H} {410, 2347}

\bibitem[\protect\citeauthoryear{{Heckman} \& {Best}}{{Heckman} \&
  {Best}}{2014}]{HB14}
{Heckman} T.~M.,  {Best} P.~N.,  2014, \mn@doi [\araa]
  {10.1146/annurev-astro-081913-035722}, \href
  {http://cdsads.u-strasbg.fr/abs/2014ARA%26A..52..589H} {52, 589}

\bibitem[\protect\citeauthoryear{{Ho}}{{Ho}}{1999}]{HO99}
{Ho} L.,  1999, in {Chakrabarti} S.~K.,  ed.,  Astrophysics and Space Science
  Library Vol. 234, Observational Evidence for the Black Holes in the Universe.
  p.~157, \mn@doi{10.1007/978-94-011-4750-7_11}

\bibitem[\protect\citeauthoryear{{Hu}}{{Hu}}{2008}]{HU08}
{Hu} J.,  2008, \mn@doi [\mnras] {10.1111/j.1365-2966.2008.13195.x}, \href
  {http://cdsads.u-strasbg.fr/abs/2008MNRAS.386.2242H} {386, 2242}

\bibitem[\protect\citeauthoryear{{Hu}}{{Hu}}{2009}]{HU09}
{Hu} J.,  2009, preprint, \href
  {http://cdsads.u-strasbg.fr/abs/2009arXiv0908.2028H} {} (\mn@eprint {arXiv}
  {0908.2028})

\bibitem[\protect\citeauthoryear{{Jahnke} \& {Macci{\`o}}}{{Jahnke} \&
  {Macci{\`o}}}{2011}]{JM11}
{Jahnke} K.,  {Macci{\`o}} A.~V.,  2011, \mn@doi [\apj]
  {10.1088/0004-637X/734/2/92}, \href
  {http://cdsads.u-strasbg.fr/abs/2011ApJ...734...92J} {734, 92}

\bibitem[\protect\citeauthoryear{{King}}{{King}}{2003}]{KI03}
{King} A.,  2003, \mn@doi [\apjl] {10.1086/379143}, \href
  {http://cdsads.u-strasbg.fr/abs/2003ApJ...596L..27K} {596, L27}

\bibitem[\protect\citeauthoryear{{Kormendy} \& {Ho}}{{Kormendy} \&
  {Ho}}{2013}]{KH13}
{Kormendy} J.,  {Ho} L.~C.,  2013, \mn@doi [\araa]
  {10.1146/annurev-astro-082708-101811}, \href
  {http://cdsads.u-strasbg.fr/abs/2013ARA%26A..51..511K} {51, 511}

\bibitem[\protect\citeauthoryear{{Kormendy} \& {Kennicutt}}{{Kormendy} \&
  {Kennicutt}}{2004}]{KK04}
{Kormendy} J.,  {Kennicutt} Jr. R.~C.,  2004, \mn@doi [\araa]
  {10.1146/annurev.astro.42.053102.134024}, \href
  {http://cdsads.u-strasbg.fr/abs/2004ARA%26A..42..603K} {42, 603}

\bibitem[\protect\citeauthoryear{{Kormendy} \& {Richstone}}{{Kormendy} \&
  {Richstone}}{1995}]{KR95}
{Kormendy} J.,  {Richstone} D.,  1995, \mn@doi [\araa]
  {10.1146/annurev.aa.33.090195.003053}, \href
  {http://cdsads.u-strasbg.fr/abs/1995ARA%26A..33..581K} {33, 581}

\bibitem[\protect\citeauthoryear{{Kormendy}, {Cornell}, {Block}, {Knapen}  \&
  {Allard}}{{Kormendy} et~al.}{2006}]{KC06}
{Kormendy} J.,  {Cornell} M.~E.,  {Block} D.~L.,  {Knapen} J.~H.,   {Allard}
  E.~L.,  2006, \mn@doi [\apj] {10.1086/501341}, \href
  {http://cdsads.u-strasbg.fr/abs/2006ApJ...642..765K} {642, 765}

\bibitem[\protect\citeauthoryear{{Kormendy}, {Fisher}, {Cornell}  \&
  {Bender}}{{Kormendy} et~al.}{2009}]{KO09}
{Kormendy} J.,  {Fisher} D.~B.,  {Cornell} M.~E.,   {Bender} R.,  2009, \mn@doi
  [\apjs] {10.1088/0067-0049/182/1/216}, \href
  {http://adsabs.harvard.edu/abs/2009ApJS..182..216K} {182, 216}

\bibitem[\protect\citeauthoryear{{L{\"a}sker}, {Ferrarese}, {van de Ven}  \&
  {Shankar}}{{L{\"a}sker} et~al.}{2014}]{LF14}
{L{\"a}sker} R.,  {Ferrarese} L.,  {van de Ven} G.,   {Shankar} F.,  2014,
  \mn@doi [\apj] {10.1088/0004-637X/780/1/70}, \href
  {http://cdsads.u-strasbg.fr/abs/2014ApJ...780...70L} {780, 70}

\bibitem[\protect\citeauthoryear{{Marconi} \& {Hunt}}{{Marconi} \&
  {Hunt}}{2003}]{MH03}
{Marconi} A.,  {Hunt} L.~K.,  2003, \mn@doi [\apjl] {10.1086/375804}, \href
  {http://cdsads.u-strasbg.fr/abs/2003ApJ...589L..21M} {589, L21}

\bibitem[\protect\citeauthoryear{{Martin-Navarro}, {Brodie}, {Romanowsky},
  {Ruiz-Lara}  \& {van de Ven}}{{Martin-Navarro} et~al.}{2018}]{IM18}
{Martin-Navarro} I.,  {Brodie} J.~P.,  {Romanowsky} A.~J.,  {Ruiz-Lara} T.,
  {van de Ven} G.,  2018, Nature, \href {http://dx.doi.org/10.1038/nature24999}
  {553, 307}

\bibitem[\protect\citeauthoryear{{Mihos} \& {Hernquist}}{{Mihos} \&
  {Hernquist}}{1994}]{MH94}
{Mihos} J.~C.,  {Hernquist} L.,  1994, \mn@doi [\apjl] {10.1086/187679}, \href
  {http://adsabs.harvard.edu/abs/1994ApJ...437L..47M} {437, L47}

\bibitem[\protect\citeauthoryear{{Minezaki} \& {Matsushita}}{{Minezaki} \&
  {Matsushita}}{2015}]{MM15}
{Minezaki} T.,  {Matsushita} K.,  2015, \mn@doi [\apj]
  {10.1088/0004-637X/802/2/98}, \href
  {http://cdsads.u-strasbg.fr/abs/2015ApJ...802...98M} {802, 98}

\bibitem[\protect\citeauthoryear{{Misgeld} \& {Hilker}}{{Misgeld} \&
  {Hilker}}{2011}]{MH11}
{Misgeld} I.,  {Hilker} M.,  2011, \mn@doi [\mnras]
  {10.1111/j.1365-2966.2011.18669.x}, \href
  {http://cdsads.u-strasbg.fr/abs/2011MNRAS.414.3699M} {414, 3699}

\bibitem[\protect\citeauthoryear{{Morganti}, {Fogasy}, {Paragi}, {Oosterloo}
  \& {Orienti}}{{Morganti} et~al.}{2013}]{MF13}
{Morganti} R.,  {Fogasy} J.,  {Paragi} Z.,  {Oosterloo} T.,   {Orienti} M.,
  2013, \mn@doi [Science] {10.1126/science.1240436}, \href
  {http://adsabs.harvard.edu/abs/2013Sci...341.1082M} {341, 1082}

\bibitem[\protect\citeauthoryear{{Nagar}, {Falcke}  \& {Wilson}}{{Nagar}
  et~al.}{2005}]{NA05}
{Nagar} N.~M.,  {Falcke} H.,   {Wilson} A.~S.,  2005, \mn@doi [\aap]
  {10.1051/0004-6361:20042277}, \href
  {http://cdsads.u-strasbg.fr/abs/2005A%26A...435..521N} {435, 521}

\bibitem[\protect\citeauthoryear{{Nyland} et~al.,}{{Nyland}
  et~al.}{2016}]{NY16}
{Nyland} K.,  et~al., 2016, \mn@doi [\mnras] {10.1093/mnras/stw391}, \href
  {http://cdsads.u-strasbg.fr/abs/2016MNRAS.458.2221N} {458, 2221}

\bibitem[\protect\citeauthoryear{{Park}, {Sohn}  \& {Yi}}{{Park}
  et~al.}{2013}]{PA13}
{Park} S.,  {Sohn} B.~W.,   {Yi} S.~K.,  2013, \mn@doi [\aap]
  {10.1051/0004-6361/201321310}, \href
  {http://cdsads.u-strasbg.fr/abs/2013A%26A...560A..80P} {560, A80}

\bibitem[\protect\citeauthoryear{Pellegrini}{Pellegrini}{2010}]{PE10}
Pellegrini S.,  2010, The Astrophysical Journal, 717, 640

\bibitem[\protect\citeauthoryear{{Peng}}{{Peng}}{2007}]{PE07}
{Peng} C.~Y.,  2007, \mn@doi [\apj] {10.1086/522774}, \href
  {http://cdsads.u-strasbg.fr/abs/2007ApJ...671.1098P} {671, 1098}

\bibitem[\protect\citeauthoryear{{Press}, {Teukolsky}, {Vetterling}  \&
  {Flannery}}{{Press} et~al.}{1992}]{PT92}
{Press} W.~H.,  {Teukolsky} S.~A.,  {Vetterling} W.~T.,   {Flannery} B.~P.,
  1992, {Numerical recipes in FORTRAN. The art of scientific computing}

\bibitem[\protect\citeauthoryear{{Sani}, {Marconi}, {Hunt}  \&
  {Risaliti}}{{Sani} et~al.}{2011}]{SA11}
{Sani} E.,  {Marconi} A.,  {Hunt} L.~K.,   {Risaliti} G.,  2011, \mn@doi
  [\mnras] {10.1111/j.1365-2966.2011.18229.x}, \href
  {http://cdsads.u-strasbg.fr/abs/2011MNRAS.413.1479S} {413, 1479}

\bibitem[\protect\citeauthoryear{{Savorgnan} \& {Graham}}{{Savorgnan} \&
  {Graham}}{2016a}]{SG16}
{Savorgnan} G.~A.~D.,  {Graham} A.~W.,  2016a, \mn@doi [\apjs]
  {10.3847/0067-0049/222/1/10}, \href
  {http://cdsads.u-strasbg.fr/abs/2016ApJS..222...10S} {222, 10}

\bibitem[\protect\citeauthoryear{{Savorgnan} \& {Graham}}{{Savorgnan} \&
  {Graham}}{2016b}]{SAGR16}
{Savorgnan} G.~A.~D.,  {Graham} A.~W.,  2016b, \mn@doi [\mnras]
  {10.1093/mnras/stv2713}, \href
  {http://cdsads.u-strasbg.fr/abs/2016MNRAS.457..320S} {457, 320}

\bibitem[\protect\citeauthoryear{{Savorgnan}, {Graham}, {Marconi}, {Sani},
  {Hunt}, {Vika}  \& {Driver}}{{Savorgnan} et~al.}{2013}]{SA13}
{Savorgnan} G.,  {Graham} A.~W.,  {Marconi} A.,  {Sani} E.,  {Hunt} L.~K.,
  {Vika} M.,   {Driver} S.~P.,  2013, \mn@doi [\mnras] {10.1093/mnras/stt1027},
  \href {http://cdsads.u-strasbg.fr/abs/2013MNRAS.434..387S} {434, 387}

\bibitem[\protect\citeauthoryear{{Silk} \& {Rees}}{{Silk} \&
  {Rees}}{1998}]{SR98}
{Silk} J.,  {Rees} M.~J.,  1998, \aap, \href
  {http://cdsads.u-strasbg.fr/abs/1998A%26A...331L...1S} {331, L1}

\bibitem[\protect\citeauthoryear{{Tremaine} et~al.,}{{Tremaine}
  et~al.}{2002}]{TR02}
{Tremaine} S.,  et~al., 2002, \mn@doi [\apj] {10.1086/341002}, \href
  {http://cdsads.u-strasbg.fr/abs/2002ApJ...574..740T} {574, 740}

\bibitem[\protect\citeauthoryear{{Trujillo}, {Graham}  \& {Caon}}{{Trujillo}
  et~al.}{2001}]{TR01}
{Trujillo} I.,  {Graham} A.~W.,   {Caon} N.,  2001, \mn@doi [\mnras]
  {10.1046/j.1365-8711.2001.04471.x}, \href
  {http://cdsads.u-strasbg.fr/abs/2001MNRAS.326..869T} {326, 869}

\makeatother
\end{thebibliography}








\bsp	
\label{lastpage}
\end{document}